\documentclass[letter,traditabstract]{aa} 
\usepackage{graphicx}
\usepackage{txfonts}
\usepackage{natbib}
\bibpunct{(}{)}{;}{a}{}{,}

\begin{document}

\title{WASP-33: The first $\delta$ Scuti exoplanet host star}

\author{E.~Herrero$^{1}$ \and 
J.C.~Morales$^{1}$ \and I.~Ribas$^{1}$ \and 
R.~Naves$^{2}$}

\institute{Institut de Ci\`{e}ncies de l'Espai (CSIC-IEEC), Campus UAB, 
Facultat de Ci\`{e}ncies, Torre C5 parell, 2a pl, 08193 Bellaterra, 
Spain, \email{eherrero@ice.cat, morales@ice.cat, 
iribas@ice.cat}\label{inst1}
\and
Observatori Montcabrer, C/Jaume Balmes, 24, Cabrils, Spain, 
\email{ramonnavesnogues@gmail.com}\label{inst2}
}
\date{Received <date> /
Accepted <date>}

\abstract{We report the discovery of photometric oscillations in the host 
star of the exoplanet WASP-33 b (HD 15082). The data were obtained in the 
$R$ band in both transit and out-of-transit phases from the 0.3-m 
telescope and the Montcabrer Observatory and the 0.8-m telescope at the 
Montsec Astronomical Observatory. Proper fitting and subsequent removal of 
the transit signal reveals stellar photometric variations with a 
semi-amplitude of about 1 mmag. The detailed analysis of the 
periodogram yields a structure of significant signals around a frequency 
of 21 cyc d$^{-1}$, which is typical of $\delta$ Scuti-type variable 
stars. An accurate study of the power spectrum reveals a possible
commensurability with the planet orbital motion with a factor of 26, but
this remains to be confirmed with additional time-series data that will
permit the identification of the significant frequencies.
These findings make WASP-33 the first
transiting exoplanet host star with $\delta$ Sct variability and a very
interesting candidate to search for star-planet interactions.}


\maketitle

\section{Introduction}

Over 100 transiting exoplanets have been confirmed to date, most of them 
orbiting solar-type or late-type stars. WASP-33 b, a gas giant planet 
showing transits on a fast-rotating main-sequence A5 star, represents an 
uncommon case that offers the possibility of studying an intermediate-mass
main-sequence star. 
WASP-33 b was first reported as a transiting planet candidate 
by \cite{2006MNRAS.372.1117C}, but it was not officially announced as an 
exoplanet until the study of \cite{2010MNRAS.407..507C}. The relatively 
long time lapse may be explained because the host star (HD 15082, $V=8.3$) 
is a fast rotator ($v\sin i=86\ \rm km s^{-1}$) and this hampers precise 
radial velocity work. \cite{2010MNRAS.407..507C} carried out a detailed 
study considering both photometry and spectral line profile variations 
during transits and established an upper mass limit of $4.1\ \rm M_{J}$ 
for the planet. In addition, the authors presented evidence of non-radial 
pulsations in the star and suggested $\gamma$ Dor-type variability. 
Furthermore, they also found that WASP-33 b orbits the star in retrograde 
motion and that the orbit is inclined relative to the stellar equator.

Here we use observations taken at the amateur Montcabrer Observatory 
and the professional fully robotic Montsec Astronomical Observatory -- 
OAdM \citep{2008SPIE.7019..192K}, as well as additional observations from 
the Exoplanet Transit Database (hereafter ETD, http://var2.astro.cz/ETD). 
These allow us to provide the first evidence for photometric oscillations 
on the star WASP-33, and to analyse their amplitude and periodicity. The 
presence of a large planet close to a star may cause tidal effects 
that are responsible for multiperiodic non-radial pulsations, and in special cases 
radial pulsations, on its host star \citep{2010AN....331..489S}. However, 
there are only a few known exoplanets orbiting pulsating stars, such as V391 Pegasi 
(sdB type), whose planet was discovered using the timing method 
\citep{2007Natur.449..189S}. \cite{2005A&A...440..615B} performed the 
first asteroseismic analysis for a solar-like planet host star, $\mu$ 
Arae, modelling 43 p-modes which were previously identified by 
\cite{2005A&A...440..609B}. Recently, Kepler data were used to 
characterize the exoplanet host HAT-P-7 through an analysis of its 
simultaneously discovered solar-like oscillations 
\citep{2010ApJ...713L.164C}. WASP-33 is the first case where $\delta$ 
Sct pulsations have been observed in a known transiting exoplanet host 
star.

\section{Observations and photometry}

The first observations in our dataset were obtained from Montcabrer 
Observatory on 2010 August 26, September 7 and 14 and October 20,
using a 0.3-m Schmidt-Cassegrain telescope and a SBIG ST8-XME camera 
with an AO-8 adaptive optics system, working at a 1.03\arcsec per pixel 
scale. This is an amateur private observatory with the Minor Planet Center 
code 213, which is located in the suburbs of Barcelona, Spain. Photometry on two 
additional nights with very good photometric conditions (September 21 and 
28, 2010) was obtained from OAdM and with a fully robotic 80-cm 
Ritchey-Chr\'etien telescope and a FLI PL4240 2k$\times$2k camera with a 
plate scale of 0.36\arcsec per pixel. All photometry described above 
was carried out in the Johnson $R$ band and by defocusing the images to 
increase the photometric precision. Aperture photometry was performed on 
the images, using three or four comparison stars, as available. The first 
transit was observed with the aim of improving the definition of the 
existing light curves, most of them incomplete so far 
\citep{2010MNRAS.407..507C}, whereas the follow-up photometry was obtained 
at several orbital phases in order to further study the oscillations that 
the first photometric dataset seemed to suggest.

Additional recent transit photometry of WASP-33 is available at the ETD, 
including some diagrams and a parameter analysis of each transit. WASP-33 
b transit $R$-band photometric data from K. Hose (August 23, 
2010), F. Hormuth (August 26, 2010) and C. Lopresti 
(September 11, 2010) were used in our analysis. All 
photometry datasets used in this study are listed in Table~\ref{tab:data}. 
The root mean square (rms) residuals given in the third column, even though
they are affected by the pulsations, give an idea of the precision of the
photometry from each observatory.

\begin{table}
\caption{Photometric datasets used in this work. Montcabrer Observatory 
and OAdM data were specially acquired during the course of this work. The 
rest of the photometry is public at the ETD. In transit observations, the 
photometric root mean square (rms) residual per measurement and 
the transit time deviation (TTD) with respect to the mid-transit 
ephemeris given in \cite{2010MNRAS.407..507C} are calculated 
from the fits represented in Fig.~\ref{fig:transits}.}
\label{tab:data}
\centering
\begin{tabular}{ccccc}
\hline\hline
Author/    &           & Transit / & rms    & TTD  \\
Observatory& Date      & Out of tr.& (mmag) & (min)\\
\hline
K. Hose    & 23/08/2010\phantom{a} & T   & 3.7 &  $4.10\pm1.31$ \\
Montcabrer & 26/08/2010a& T   & 2.8 & $9.76\pm1.01$ \\
F. Hormuth & 26/08/2010b& T   & 2.7 & $3.18\pm0.86$ \\
Montcabrer & 07/09/2010\phantom{a} & OOT & 3.1 &  \\
C. Lopresti& 11/09/2010\phantom{a} & T   & 3.9 & $-1.66\pm1.62$ \\
Montcabrer & 14/09/2010\phantom{a} & OOT & 2.3 & \\
OAdM       & 21/09/2010\phantom{a} & OOT & 1.6 & \\
OAdM       & 28/09/2010\phantom{a} & T   & 2.1 & $2.83\pm0.50$ \\
Montcabrer & 20/10/2010\phantom{a} & T   & 1.8 & $7.55\pm0.86$ \\
\hline
\hline
\end{tabular}
\end{table}

\section{Data analysis and discussion}
\label{sec:an}

\begin{figure}
\centering
   \includegraphics[width=\columnwidth]{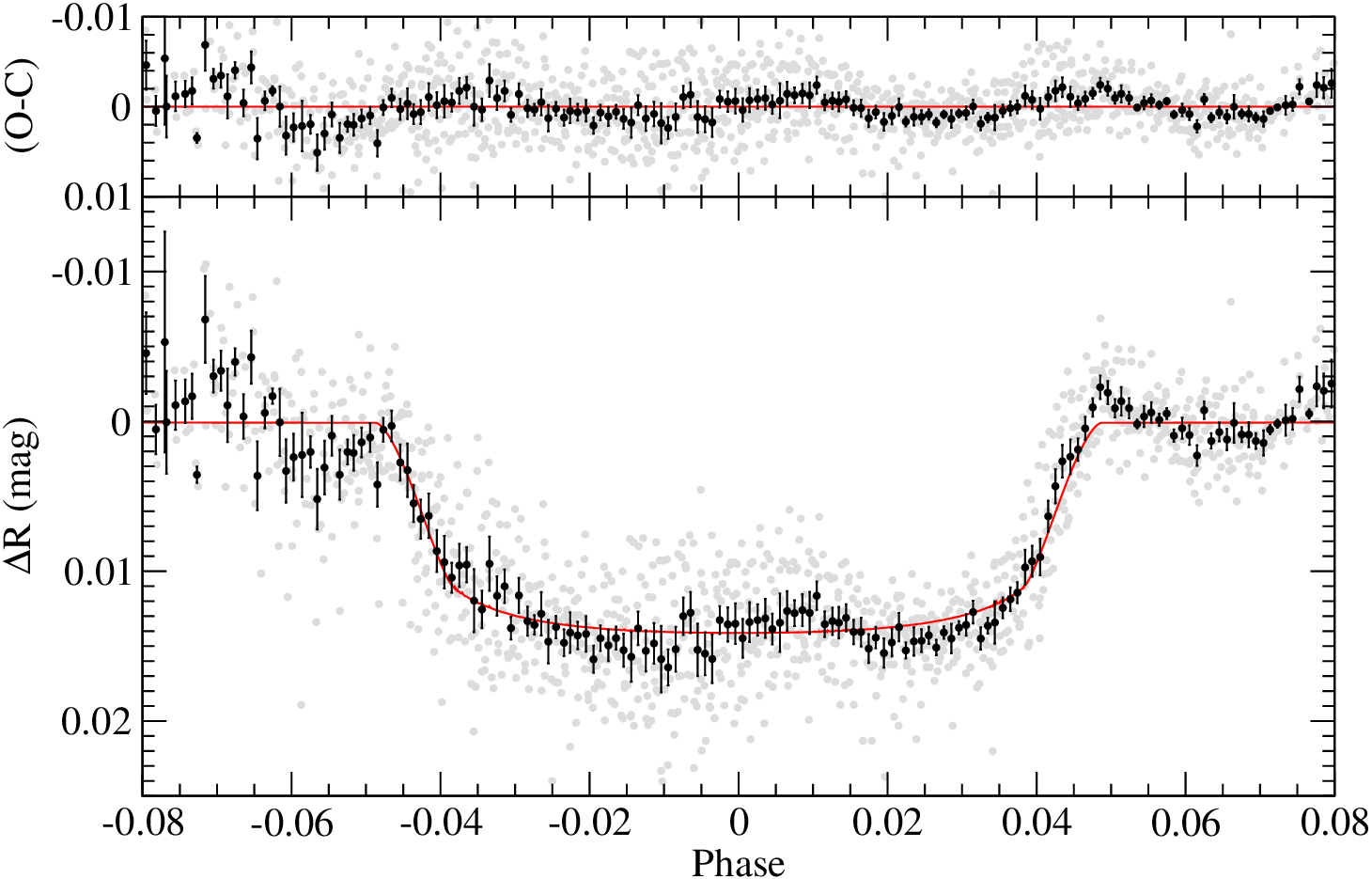}
    \caption{Best fit to the overall $R$-band transit photometry 
phase-folded (grey symbols) and binned in steps of 0.001 in phase (black 
symbols) for better visualization. The top panel shows the residuals of
this fit. Error bars are computed as the mean value error of each bin.}
     \label{fig:transit_comb}
\end{figure}

\begin{figure}
\centering
   \includegraphics[width=\columnwidth]{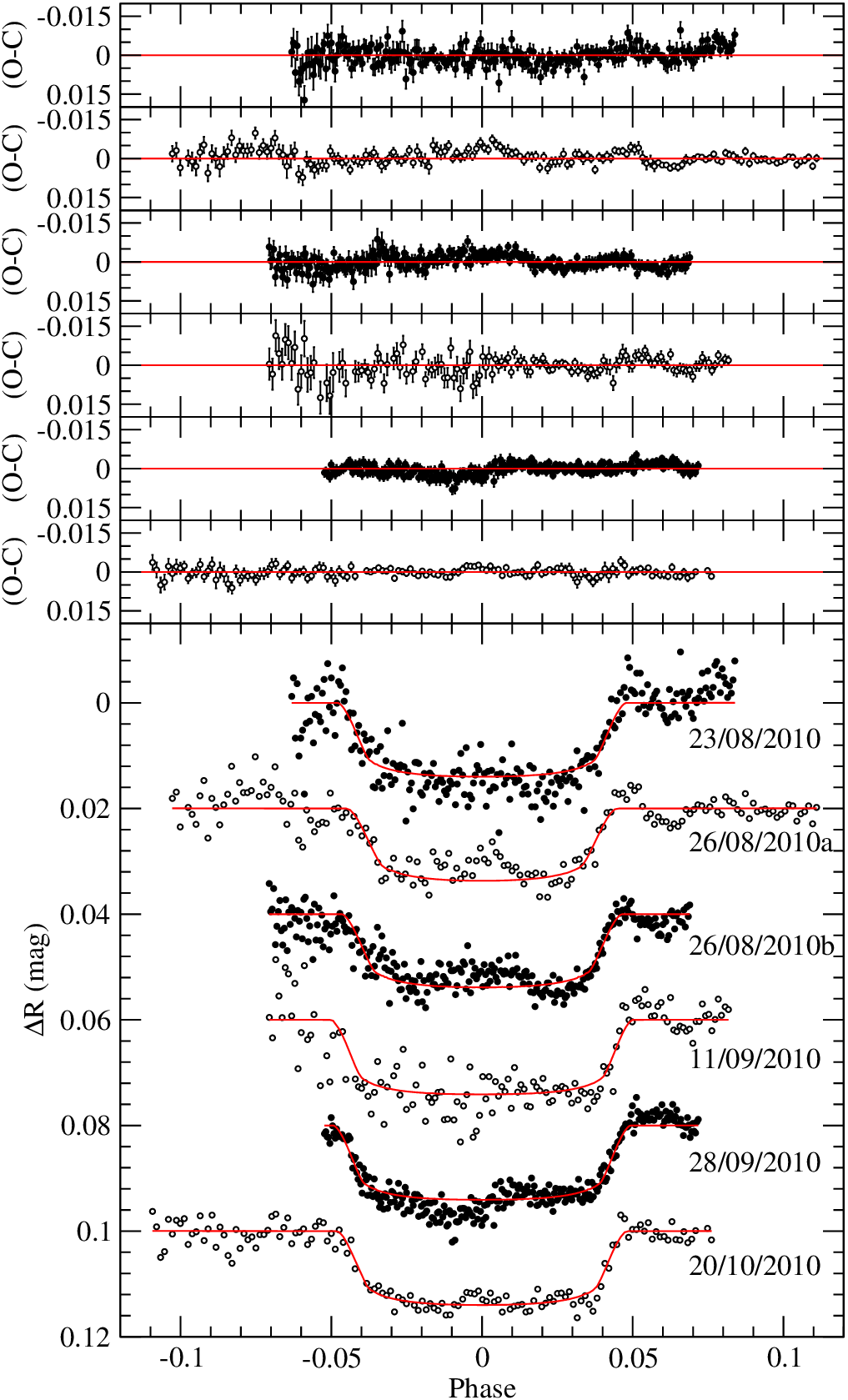}
    \caption{$R$-band transit photometry of WASP-33. The solid line is the 
best fit to each curve with the relative radii of the star and the planet
fixed to those found in the global fit using the JKTEBOP code. The 
residuals are shown in the upper panels in the same order as transits are 
displayed. See Table~\ref{tab:data} for reference.}
     \label{fig:transits}
\end{figure}

The transit photometry datasets were analysed with the JKTEBOP code 
\citep{2004MNRAS.351.1277S,2004MNRAS.355..986S}, which is based on the 
EBOP code written by Paul B. Etzel \citep{1981AJ.....86..102P}. Each 
transit light curve was corrected for a slight trend using low-order 
polynomials. Preliminary fits were run with the JKTEBOP by solving for the 
sum of the radii relative to the orbital semi-major axis (star + planet; 
$r_1 + r_2$), ratio of radii ($r_2/r_1$), inclination, and time of 
transit. This was done before combining all transits by including a 
shift to normalize their out-of-transit level, so that consistent 
parameters were obtained for all the transits. The period was fixed to 
that given in \cite{2010MNRAS.407..507C}, $P_{\rm orb}=1.2198669$ d. To
properly weigh the observations from different authors, we performed a running
average of the residuals from a preliminary fit in bins of 20 minutes, 
selecting as individual error for each point the standard deviation of its 
bin. The best fit to the overall transit photometry is shown in 
Fig.~\ref{fig:transit_comb} and the associated parameters are given in 
Table~\ref{tab:parameters}. All parameters are consistent within the 
uncertainties with those in \cite{2010MNRAS.407..507C}. A pulsation trend 
is clearly present on the residuals of this fit, and as Fig. 
\ref{fig:transits} illustrates, it is already apparent to the eye that all 
the curves (some of them separated by almost 60 d) show ``bumps'' 
that always appear at the similar orbital phases (note, e.g., the phases just 
after the egress).

\begin{table}
\caption{Transit parameters fitted using JKTEBOP. The transit depth was
computed from the fit. The error bars are 1-$\sigma$ uncertainties.}
\label{tab:parameters}
\centering
\begin{tabular}{cc}
\hline\hline
Parameter               & Value                  \\
\hline
$r_{1}+r_{2}$           & 0.309 $\pm$ 0.007      \\
$r_{2}/r_{1}$           & 0.1066 $\pm$ 0.0008    \\
$i$ ($^{\circ}$)        & 85.8 $\pm$ 2.0         \\
$T_{0}-2450000$         & 5431.8879 $\pm$ 0.0003 \\
rms (mmag)              & 3.44                   \\                                                        
Depth (mmag)            & 14.3 $\pm$ 0.2         \\
\hline
\end{tabular}
\end{table}

\begin{figure}
\centering
   \includegraphics[width=\columnwidth]{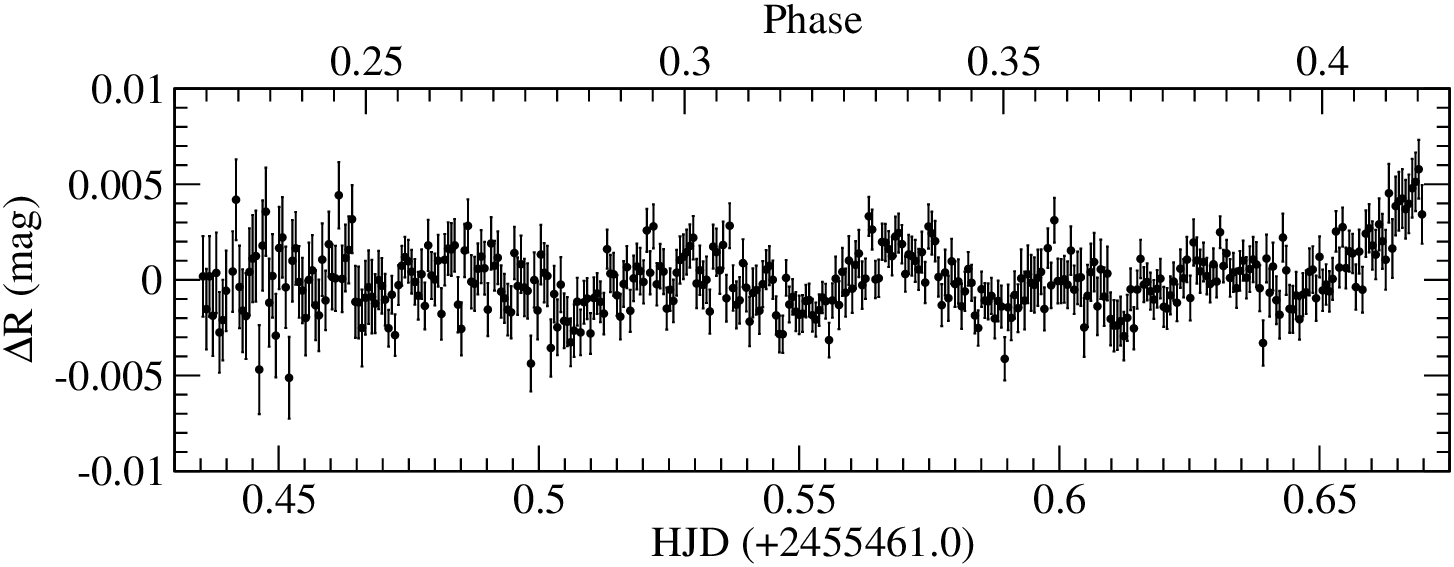}
    \caption{Out-of-transit photometry of WASP-33 obtained from OAdM on
21 September, 2010.}
     \label{fig:oot_lc}
\end{figure}

To better determine the characteristic period of the observed 
pulsations, we carried out individual fits to each transit by leaving the 
inclination and transit time as free parameters and fixing the relative 
radii of the star and the planet according to the global result. This was 
done to minimize the possible effects of transit time variations caused by 
unknown third bodies in the system on the residuals and the effect of 
trend corrections on the transit depth. We find that the inclination of 
each individual fit is well within the uncertainty of that determined for 
the combination of the transits. In Table~\ref{tab:data} the transit time
deviations of each transit are also listed, however, their scarcity and
possible systematics caused by the pulsation precludes us of drawing any
conclusions about the presence of additional planets in the system.

\begin{figure}
\centering
   \includegraphics[width=\columnwidth]{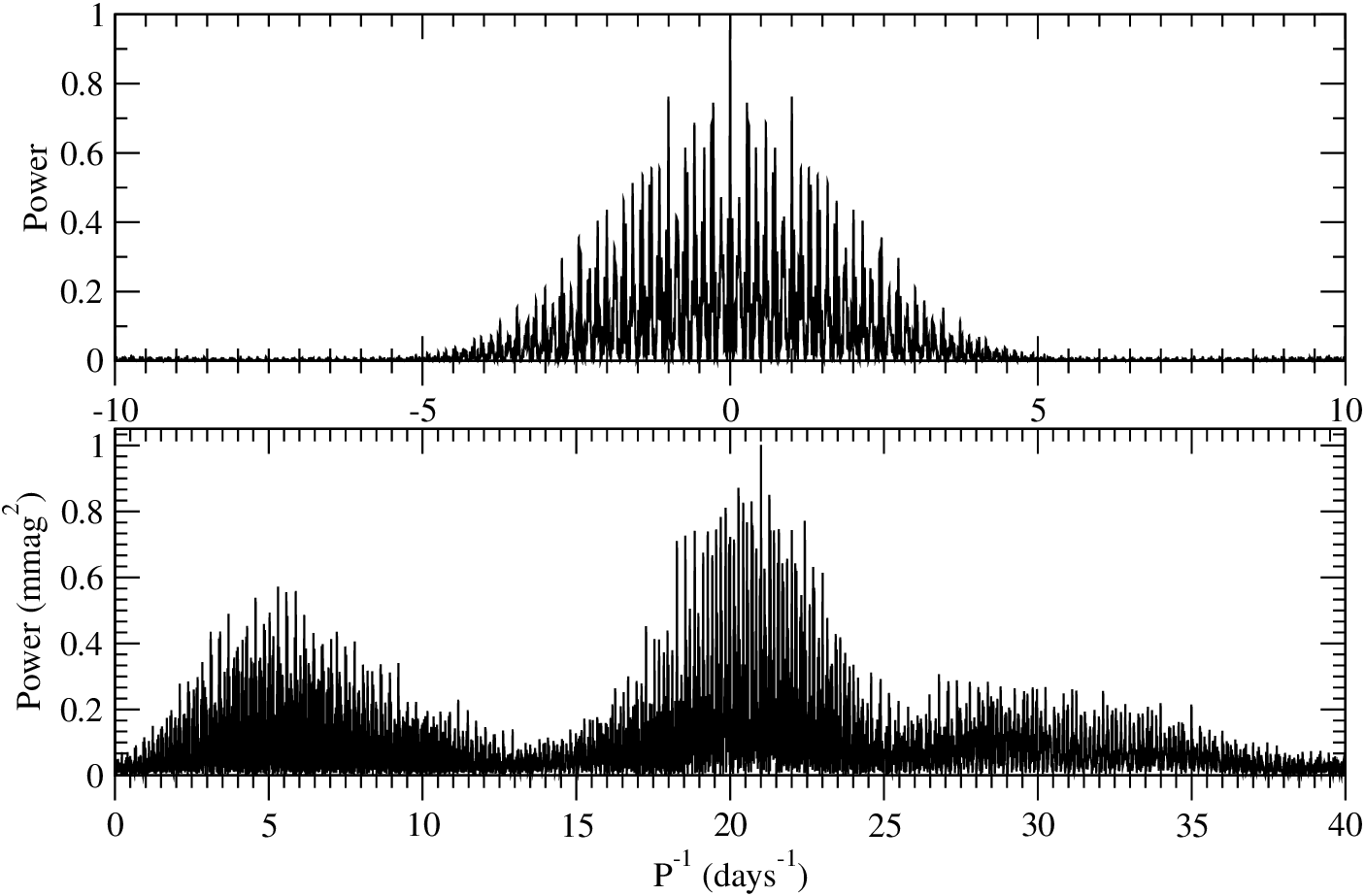}
     \caption{Periodogram of the transit residuals and out-of-eclipse 
phases of WASP-33 using PERIOD04 code. Top panel displays the
window function.}
     \label{fig:period}
\end{figure}

As already shown, the residuals of the transit fits show a clear 
oscillation pattern that is also present in the out-of-transit photometry 
obtained for WASP-33. The presence of oscillations during the entire 
orbital period, including out-of-transit phases as shown in
Fig.~\ref{fig:oot_lc}, rules out the possibility of the ``bumps'' seen in transit 
as being caused by the effect of the gravity darkening in a rapidly 
rotating star such as WASP-33 \citep{2009ApJ...705..683B}. Thus, we 
combined the residuals from the transit fits and the out-of-transit light 
curves in a single curve to better determine the period of the 
oscillations. Figure~\ref{fig:period} shows the results of running a 
periodogram on these data with the 
PERIOD04 code \citep[][http://www.univie.ac.at/tops/period04/]{2005CoAst.146...53L}
that is based on the discrete Fourier 
transform algorithm. A main peak around a frequency of 21 d$^{-1}$ is 
clearly present in this figure, corresponding to a period of about 68~min. 
However, a close inspection of this peak reveals that it is composed of 
several other peaks that are an imprint of the window function associated 
to our data. The second structure around 6~d$^{-1}$, which is similar
to the typical length of the individual photometric series, is probably spurious
and may be related to the detrending and the flux-shift of
the different light curves.

\begin{figure}
\centering
   \includegraphics[width=\columnwidth]{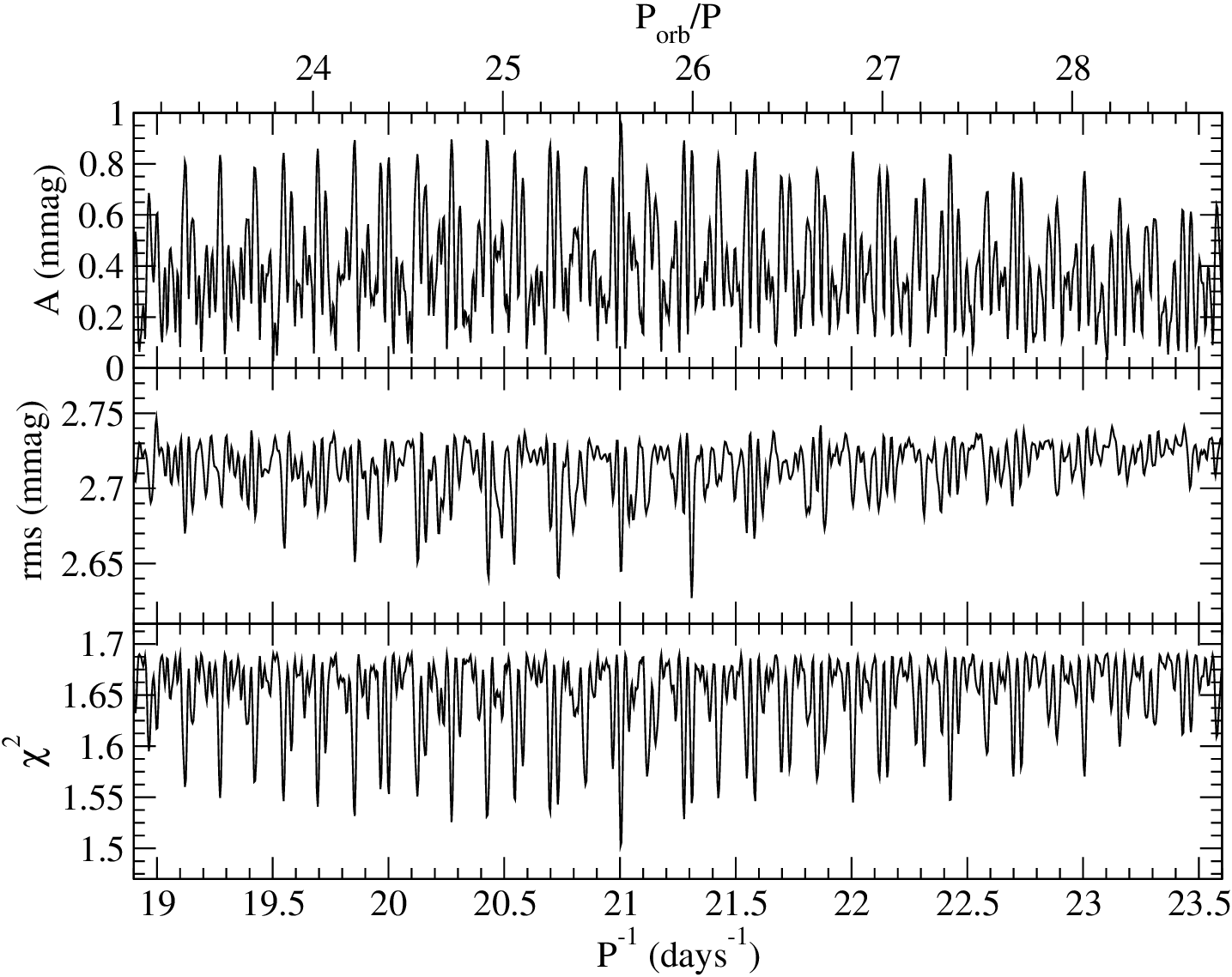}
     \caption{Results from the Levenberg-Marquardt algorithm method displaying
the structure of the main periodogram peak. Both amplitude and $\chi^{2}$ give
best fits for 21.00~d$^{-1}$, while the rms is minimized for a pulsation of
frequency 21.31~d$^{-1}$, which is commensurable with the planet's orbital
period.}
     \label{fig:periodLM}
\end{figure}

\begin{figure}
\centering
   \includegraphics[width=\columnwidth]{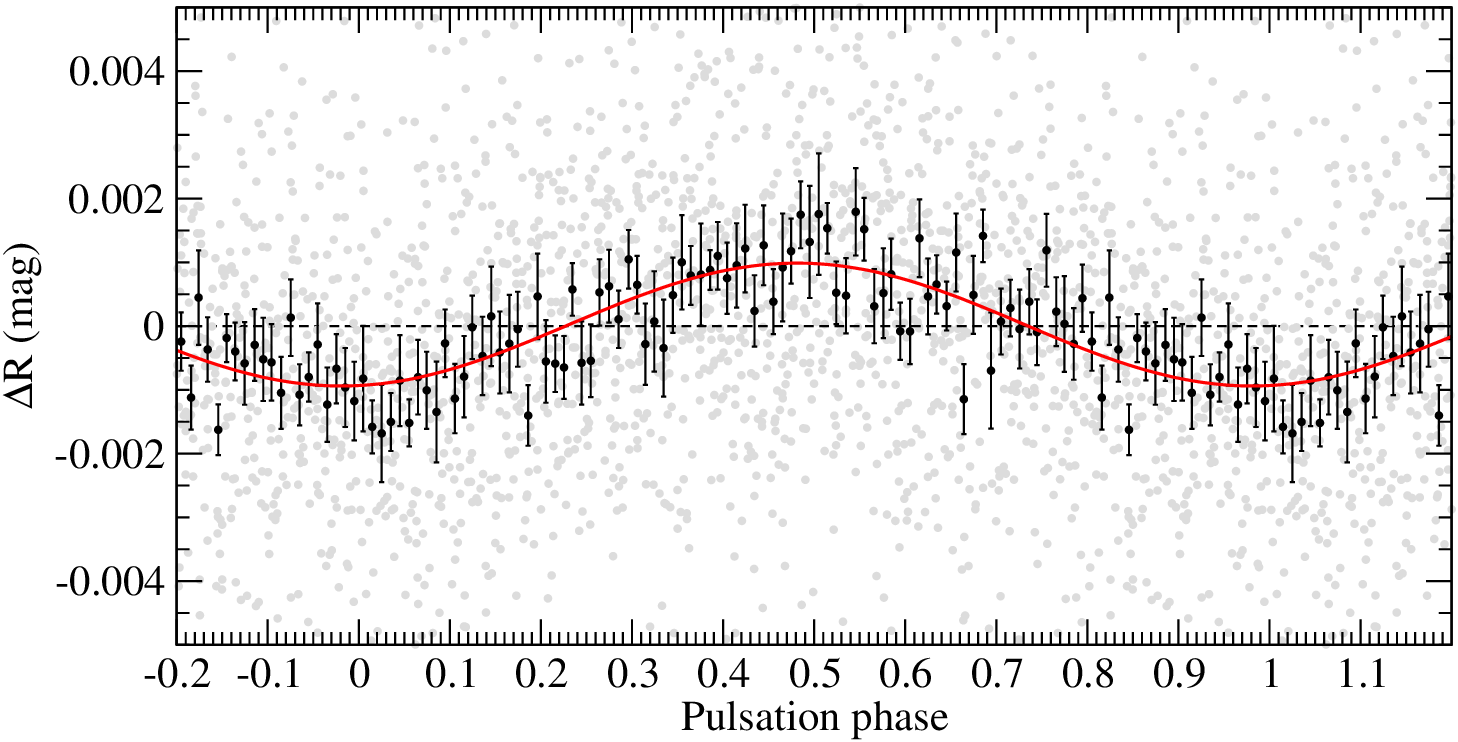}
     \caption{Residuals of the transit fits and out-of-eclipse photometry 
of WASP-33 phase-folded according to the period of the pulsation found in 
this work, i.e., 68.56$\pm$0.02~min, (grey symbols) and 0.01 phase binning 
(black symbols). The solid curve is the best fit sinusoidal modulation 
with an amplitude of 0.98~mmag.}
     \label{fig:pulsation}
\end{figure}

To better analyse the structure of this peak, we fitted the 
theoretical transit light curve to the original data by adding a 
sinusoidal modulation of the light of the system, thus performing a phase
dispersion minimization algorithm \citep{1978ApJ...224..953S}.
The minimization of this parametric fit was performed by means of a
Levenberg-Marquardt algorithm \citep{1992nrfa373.231}. 
Because of the degeneracy on the solutions, we fitted the amplitude of the 
modulation and tested a grid of frequency and phase difference values.
A key element to consider is the weighing of each data point. At the high
precision of our photometry, photon noise is not a good 
measure of the uncertainty of the measurement because systematic effects in the
form of correlated noise dominate \cite[e.g.,][]{2006MNRAS.373.231}. Therefore,
we applied a two-step process in which we performed the 
analysis assuming an initial constant arbitrary value of the standard
deviation $\sigma$=1 for each data point. The best-fitting solution
was then used to calculate a new value for $\sigma$ as the local value of the
rms residual (computed as the running average of the residuals in bins of 20 minutes),
and a final fit was run with weights set accordingly. 
As can be seen in Fig.~\ref{fig:periodLM}, the lowest $\chi^{2}$ is found for a
pulsation with a semi-amplitude of 0.98$\pm$0.05~mmag and a frequency of
21.004$\pm$0.004~d$^{-1}$ at a phase difference with the mid-transit time of 
264$\pm$12$^\circ$ (at reference epoch from Table \ref{tab:parameters}).
This corresponds to a period of
$P=68.56\pm0.02$~min. Figure~\ref{fig:pulsation} shows the overall
residuals and the out-of-eclipse photometry phase-folded according to this
best-fitting period. As a result of the separation between observing nights,
the main feature in Fig.~\ref{fig:periodLM} is composed of many equidistant
peaks with similar $\chi^{2}$. Interestingly, the periodogram peak with lower
resulting rms differs from the one above as it occurs for a pulsation of about
0.86~mmag in semi-amplitude with frequency of 21.311$\pm$0.004~d$^{-1}$
corresponding to a pulsation period of $67.57\pm0.02$~min. This illustrates the
relatively strong impact of the weighing scheme because the signal is weak and our
time series is relatively short. The latter frequency is especially interesting
as is happens to be commensurable with the planet's orbital motion with a
factor of 25.997$\pm$0.005.

Line-profile tomography in \cite{2010MNRAS.407..507C} provided strong 
evidence for non-radial pulsations in WASP-33 with a period around one 
day, similar to those usually found in $\gamma$ Dor stars. Moreover, the 
authors point out the possibility that the retrograde orbiting planet 
could be tidally inducing them. Both the period of the photometric 
oscillations presented here and the stellar properties from 
\cite{2010MNRAS.407..507C} ($T_{\rm eff}=7400\pm200\ \rm K$, $\log 
g=4.3\pm0.2$) locate WASP-33 well within the $\delta$ Sct instability 
strip.  \cite{2002MNRAS.333..251H} present a discussion on the different 
properties of $\delta$ Sct and $\gamma$ Dor pulsators. Using the formalism 
there it can be shown that the pulsation constant of WASP-33 ($\log Q_{\rm 
WASP-33}=-1.45$) perfectly corresponds to the $\delta$ Sct domain. The 
power spectrum of $\delta$ Sct pulsators is usually very rich, as 
illustrated by the 75+ frequencies identified for FG Vir 
\citep{2005A&A...435..955B}, but asteroseismic modelling is especially 
difficult for fast rotators as WASP-33. Given this 
evidence, a different scenario considering that WASP-33 belongs to the 
relatively rare class of hybrid pulsators, showing simultaneous $\delta$ 
Sct and $\gamma$ Dor oscillations
\citep{2002MNRAS.333..262H,2009MNRAS.398.1339H},
may be possible.

One scenario that should be explored in spite of all the evidence is the 
possibility that the photometric variations are ellipsoidal in nature and 
are not caused by pulsations. Indeed, the tidal bulge travels particularly fast 
over the stellar surface, because the orbital motion of the planet is retrograde 
with respect to the stellar rotation. Note that there is an indeterminacy 
in the stellar rotation velocity because it is not a given assumption that 
the inclination of the stellar spin axis corresponds to the planet's 
orbital inclination. But a simple calculation, equalling the 
pulsation frequency found here to that of the relative orbital motion of 
WASP-33 b above the star surface, renders the ellipsoidal variation 
scenario as physically not valid since the star would have to rotate at a 
largely super-critical speed in terms of gravitational break-up. It is 
more likely that the tidal bulge rotates over the stellar surface with a 
frequency of about 2~d$^{-1}$, which is the net combination of the 
orbital and rotation frequencies, and is far from the frequencies we find 
relevant in the periodogram.

As seen above, the periodogram is still not defined sufficiently
well to assess the true pulsation spectrum of WASP-33. However, given the
potential interest, we find it appropriate to briefly address here the
possibility of the coupling between the pulsation and the orbit. If the
pulsation frequency of 21.311$\pm$0.004~d$^{-1}$ turned out to be real, this
could suggest the existence of some kind of star-planet interactions that lead
to a high-order commensurability of 26 between the pulsation and 
orbital periods. Very few cases are known to date of a $\delta$ Sct variable
belonging to a close binary system \citep{2005ASPC..333...52W}. The best
described candidate for tidally-induced oscillations is HD 209295, which 
simultaneously presents $\gamma$ Dor and $\delta$ Sct-type pulsations, and 
which was photometrically found to show several p-modes directly related 
to the orbital motion of its companion \citep{2002MNRAS.333..262H}. This poses
the tantalizing possibility of a similar situation occurring in
WASP-33, which will certainly be much clearer when additional photometry
is acquired and puts constraints on the $\delta$ Sct frequency spectrum.

\section{Conclusions}

High-precision $R$-band photometry has allowed us to present the discovery 
of photometric oscillations in WASP-33, thus becoming the first transiting
planet host with $\delta$ Sct pulsations. These oscillations have a
period of about 68 minutes and a semi-amplitude of 1~mmag. The frequency
spectrum of WASP-33 is still not well defined because the size and significance of
the peaks depends on the scheme used to weigh the data. In a
particular weighing criterion, the most significant period comes out to be
commensurable with the orbital period at a factor of 26.
If this
association is assumed to be physically relevant, the multiplicity hints 
at the existence of planet-star interactions. 
In any case, gaining insight into the nature of WASP-33  
will necessarily require the collection of 
additional data (possibly multi-colour) with longer time baseline and to 
carry out pulsation and dynamical modelling.
The WASP-33 system now represents a new benchmark in the world of 
exoplanets that can provide valuable information on stellar pulsations 
(through, e.g., transit surface mapping), on the tidal interactions 
between planets and stars, and on the dynamical evolution of planetary 
systems.

\acknowledgements
We thank G. Handler, P.~J. Amado and C. Jordi for useful discussions. The 
referee, M. Bazot, is thanked for insightful comments that led to the
improvement of the paper. Data were partially obtained by J. Martin with
the Joan Or\'o Telescope (TJO) of the Montsec Astronomical Observatory
(OAdM), which is owned by the Consorci del Montsec and operated by the
Institute for Space Studies of Catalonia (IEEC). This research has made
use of the Exoplanet Transit Database (http://var.astro.cz/ETD).

\bibliographystyle{aa} 

\end{document}